# On the importance of low-frequency signals in functional and molecular photoacoustic computed tomography

Tri Vu, Paul Klippel, Aidan J. Canning, Chenshuo Ma, Huijuan Zhang, Ludmila A. Kasatkina, Yuqi Tang, Jun Xia, Vladislav V. Verkhusha, Tuan Vo-Dinh, Yun Jing, and Junjie Yao

***Abstract*—In photoacoustic computed tomography (PACT) with short-pulsed laser excitation, wideband acoustic signals are generated in biological tissues with frequencies related to the effective shapes and sizes of the optically absorbing targets. Low-frequency photoacoustic signal components correspond to slowly varying spatial features and are often omitted during imaging due to the limited detection bandwidth of the ultrasound transducer, or during image reconstruction as undesired background that degrades image contrast. Here we demonstrate that low-frequency photoacoustic signals, in fact, contain functional and molecular information, and can be used to enhance structural visibility, improve quantitative accuracy, and reduce spare-sampling artifacts. We provide an in-depth theoretical analysis of low-frequency signals in PACT, and experimentally evaluate their impact on several representative PACT applications, such as mapping temperature in photothermal treatment, measuring blood oxygenation in a hypoxia challenge, and detecting photoswitchable molecular probes in deep organs. Our results strongly suggest that low-frequency signals are important for functional and molecular PACT.

## I. INTRODUCTION

PHOTOACOUSTIC (PA) computed tomography (PACT) has developed rapidly with increasing preclinical and clinical applications over the past decades, largely due to its balanced high resolution, large penetration depth, and intrinsic sensitivity to functional and molecular contrast [1, 2]. In PACT, short-pulsed wide-field laser excitation is used for generating ultrasound waves by optically absorbing targets via the photoacoustic effect. The propagating waves can be detected by an ultrasonic transducer array to reconstruct the original optical energy deposition inside the targets. Optical-absorption contrast images are digitally reconstructed by applying inverse algorithms [3, 4]. The unique sensitivity of PACT to optical absorption contrast enables quantitative imaging of a wide range of functional and molecular properties of biological tissues including blood oxygenation [5], tissue temperature [6], and molecular probe distribution [7], usually with multispectral excitation [8, 9].

In PACT, short-pulsed (usually several nanoseconds) laser excitation of tissues results in broadband PA signals [10], with frequency components that depend on the effective target size and shape, as well as on the imaging depth [11]. Recent PACT studies have been focused on detecting high-frequency signal components to improve spatial resolution. For example, localization-based PACT methods detect high-frequency signals from microscopic PA absorbers (*e.g.*, dyed droplets and microspheres) to achieve super-resolution imaging [12, 13]. Singular value decomposition analysis in PACT highlights high-frequency signals with small singular values to identify fluctuations such as hemorrhages and microbubbles [14-16]. Low-frequency PA signal components (<1 MHz), on the other hand, are often neglected because they correspond to slowly varying spatial features or large homogenous targets, and manifest as low spatial resolution and low image contrast. Low-frequency PA signals are also suppressed due to hardware limitations and image processing. Typical PACT systems use piezoelectric ultrasonic transducers [17, 18]. These transducers usually have low detection efficiency at frequencies of <2 MHz [17-19]. Large PA targets such as the superior vena cava usually manifest as hollow structures with sharp edges without weak lumen, which is attributed to both limited detection bandwidth and optical fluence attenuation. In addition, filtered back-projection, a commonly used image reconstruction method,

This work was partially sponsored by National Institutes of Health (NIH) grants R21EB027981, R21 EB027304, RF1 NS115581 (BRAIN Initiative), R01 NS111039, R01 EB028143, R21 EB027981, R01 EB031629 and R35 GM122567; National Science Foundation (NSF) CAREER award 2144788; and Chan Zuckerberg Initiative Grant 2020-226178.
  Tri Vu, Chenshuo Ma, Yuqi Tang, and Junjie Yao are with the Photoacoustic Imaging Laboratory, Duke University, Durham, NC 27708 USA.
  Paul Klippel and Yun Jing are with the Graduate Program in Acoustics, Penn State University, University Park, PA 16802.
  Aidan Canning and Tuan Vo-Dinh are with the Department of Biomedical Engineering, Department of Chemistry, and Fitzpatrick Institute of Photonics, Duke University, Durham, NC 27708.
  Huijuan Zhang and Jun Xia are with the Department of Biomedical Engineering, State University of New York, Buffalo, NY 14260.
  Ludmila A. Kasatkina and Vladislav V. Verkhusha are with the Department of Genetics and Gruss-Lipper Biophotonics Center, Albert Einstein College of Medicine, Bronx, NY 10461.
  Tri Vu, Paul Klippel and Aidan Canning contributed equally.

often uses a ramp filter with a Hamming or Hanning window to suppress high-frequency noise and background signals [20-27]. These filters may remove low-frequency signals from large structures during image reconstruction [28].

Recently PACT has started to take advantage of the wideband PA signals by combining multi-band detection [29, 30]. In this work, we believe that low-frequency signals benefit quantitative measurements in functional and molecular PACT. First, due to the frequency-dependent acoustic attenuation, low-frequency signals penetrate deeper inside tissue [4, 31]. Second, low-frequency signals are less sensitive to speed-of-sound inhomogeneities, especially for transcranial imaging [32-34]. Last, PACT systems that detect low-frequency signals can be more cost-effective, with reduced signal sampling frequency, decreased number of detection channels, and long-pulse laser diodes [35, 36]. We believe that to better assess low-frequency signals in PACT is crucial for improving instrument design and quantitative data analysis.

Here we provided an in-depth analysis of low-frequency signals in PACT and developed a numerical method to determine the optimal cutoff frequency for frequency separation, based on the maximal contrast of the region of interest. We experimentally investigated the impact of low-frequency signal components in several representative functional and molecular PACT applications: temperature mapping in photothermal treatment using nanoparticles, molecular imaging of photoswitchable phytochromes in deep organs, and deep tissue blood oxygenation measurements in hypoxia conditions. Our experimental results show that low-frequency signal components are crucial for improving quantification accuracy in functional and molecular PACT. Ultimately, we expect this study will highlight the low-frequency PA signals for optimizing system configurations and data analysis in functional and molecular PACT.

## II. METHODS

### A. Frequency characteristics of PA signals

The frequency spectrum of time-resolved PA signals closely relates to the target's characteristic spatial dimension (the dimension of the structure of interest or the decay constant of the optical energy deposition, whichever is smaller). The wave equation describing the PA pressure field $p(\mathbf{r}, t)$ at location $\mathbf{r}$ and time $t$ upon short-pulsed laser excitation is given by [31, 37]

$$\nabla^2 c^2 p(\mathbf{r}, t) - \frac{\partial^2}{\partial t^2} p(\mathbf{r}, t) = -\Gamma A(\mathbf{r}) \frac{\partial I_e(t)}{\partial t} \quad (1)$$

where $c$ is the speed of sound, $\Gamma$ is the Grueneisen parameter, $A(\mathbf{r})$ is the volumetric density of the locally absorbed optical energy, and $I_e(t)$ is the temporal profile of the wide-field laser intensity. The PA pressure field in the frequency domain is given by

$$P(\mathbf{r}, f) = \int p(\mathbf{r}, t) e^{-j2\pi f t} dt \quad (2)$$

where $f$ is the temporal frequency. Assuming that the laser pulse is a Dirac delta pulse $I_e(t) = \delta(t)$, combining Eqs. (1) and (2) give [31]

$$\nabla^2 P(\mathbf{r}, f) + k^2 P(\mathbf{r}, f) = \frac{jk\Gamma}{c} A(\mathbf{r}, f) \quad (3)$$

where $k = \frac{2\pi f}{c}$ is the wavenumber. Solution of Eq. (3) for $P(\mathbf{r}, f)$ in a region-of-interest $V$ with three-dimensional coordinates $x, y, z$ can be approximated as [31]

$$P(\mathbf{r}, f) \approx \frac{jk\Gamma e^{-jkr_0}}{4\pi r_0 c} \iiint_V A(\mathbf{r}) e^{-jk(x+y+z)} dxdydz \quad (4)$$

where $r_0$ is the distance between the target and the detector.

From Eq. (4), the normalized power spectrum of the PA signal is given by [31]

$$S(f) = \frac{1}{\Phi_c^2(f)} \iiint R(\Delta \mathbf{r}) e^{jk(\Delta x + \Delta y + \Delta z)} d\Delta x \Delta y \Delta z \quad (5)$$

where $\Phi_c(f)$ is the transducer's frequency response, and $R(\Delta \mathbf{r})$ is the autocorrelation function of $A(\mathbf{r})$.

From Eqs. (4) and (5), we can derive two important properties of the PA signal spectrum that are relevant to this study: (1) The frequency spectrum depends on the target's characteristic dimension [38]. PA signals from a smaller target have a broader bandwidth and higher central frequency, and vice versa [11, 22, 38]. In other words, low-frequency signal components correspond to slowly-varying features of the target, such as the overall liver geometry or tumor volume. High-frequency signal components, on the other hand, represent fast-varying features inside the target, such as individual blood vessels or tumor-tissue boundaries. In many applications, functional parameters of biological tissues such as temperature vary slowly in space. Thus, low-frequency signal components are useful to quantify functional parameters at the macroscopic level. (2) With sufficient light illumination, the amplitude of the PA signal typically increases with the target size or the spatial resolution voxel size, whichever is smaller, due to the increased number of effective absorbers [22]. Assuming that the thermal noise in PA is broadband, low-frequency signal components also contain less noise than high-frequency components, providing a better signal-to-noise ratio (SNR). This is particularly important for deep-tissue applications, in which the SNR decays exponentially with the imaging depth due to the optical attenuation.

### B. Ring array-based PACT system

The ring array-based PACT system is depicted schematically in Fig. 1a. The system uses a customized full-ring-shaped transducer array (Imasonics, Inc.) combing two half-ring transducers, with 512 elements in total, 8-cm diameter, 5-MHz center frequency, and >100% receiving bandwidth optimized for the receiving mode. The ring array is connected to four pre-amplifiers (LEGION AMP, PhotoSound, Inc.) and is multiplexed with a 256-channel data acquisition system (Vantage 256, Verasonics, Inc.). The ring array is capable of acquiring a cross-sectional image with two laser pulses. For light illumination, laser pulses (1064-nm wavelength, 10-ns pulse width, and 10-Hz pulse repetition rate (PRR)) are delivered to the target through two four-branched fiber bundles (Dolan-Jenner Industries, Inc.) (Figs. 1a and 1e). The fiber outlets are positioned around the ring for uniform illumination. Laser firing and ultrasound detection are synchronized by a LabVIEW-based FPGA module (myRIO-1900, National Instrument, Inc.).

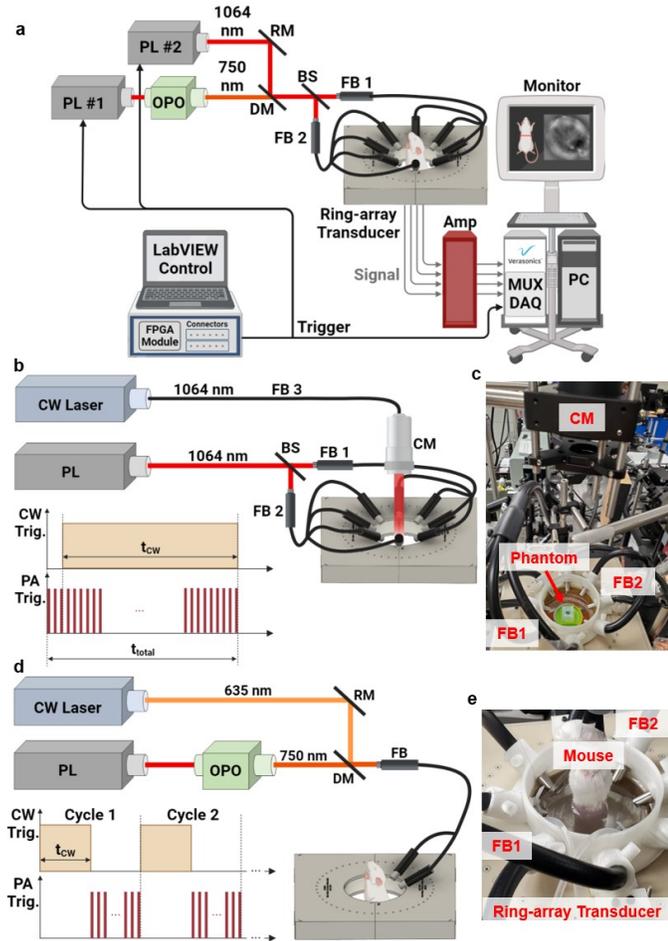

experiment. (a) Dual-wavelength whole-body illumination for hypoxia challenge experiments, showing the ring-array transducer. (b) 1064-nm light for PA illumination and heating in photothermal experiments. (c) Photograph of the photothermal experiment setup. (d) Photoswitching experiment with a 750-nm pulsed laser for PA excitation, and a 635-nm CW laser for turning on the phytochromes. (e) Photograph of the *in vivo* experiment setup with whole-body illumination. Amp, amplifier; BS, beamsplitter; DAQ, data-acquisition instrument; FB, fiber bundle; MUX, multiplexing; RM, reflection mirror; DM, dichroic mirror; CM, collimator; PL, pulsed laser.

The laser light illumination was tailored in each experiment. For photothermal treatment, continuous-wave (CW) 1064-nm light (650 mW/cm$^2$) was delivered to the heating area as illustrated in Figs. 1b and 1c. In the single-cycle treatment, the heating light was turned on for 10 minutes ($t_{cw} = 10$) following 1-minute baseline ($t_{total} = 11$). In the three-cycle treatment, a one-minute heating phase was followed by a three-minute cooling phase in each cycle. For photoswitching experiments, PA excitation and photoswitching light were instead delivered through one fiber bundle with two outlets to the kidney region of a mouse to maximize laser energy density delivered to this photoswitchable organ (Fig. 1d). A total of 11 photoswitching cycles were used. In each switching cycle, a 635-nm CW laser was turned on for the first $t_{CW} = 8$ seconds to switch on the phytochromes. For generating PA signals and switching off the phytochromes, we turned on a 750-nm pulsed OPO laser (10-ns pulse width and 10-Hz pulse repetition rate) for another 8 seconds. Before photoswitching, an anatomic image of the mouse's kidney region was acquired with ring illumination at 1064 nm. In the hypoxia challenge experiments, PA signals at 1064 nm and 750 nm were acquired around the liver region for multispectral measurement of blood oxygenation (Fig. 1e) [39]. The laser energy levels in all experiments are listed in Table I.

### C. Characterization of the system's electrical impulse response

To characterize the frequency response of the detection system, we used short-pulsed 1064-nm laser excitation on an optically-thin planar target, which generated wideband unipolar PA signals close to a delta pulse [40, 41]. The target was a thin sheet (~83-μm thickness) made of gold nanoparticles sandwiched between two layers of agarose (Figs. 2a, 2b). More information on the nanoparticle is provided in sec. II. E. The experimental setup for measuring the impulse response is depicted in Fig. 2c. The PA signal from the planar target received by the middle ring array element was used as the receiving electrical impulse response (EIR), while its frequency spectrum was the transfer function (TF) of the detection system. The receiving EIR was validated with an analytical solution and was used for Wiener deconvolution of the RF data before the image reconstruction [42].

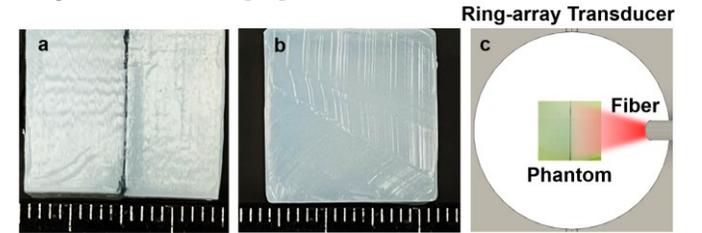

Fig. 2. Measurement of the system's receiving EIR using a planar target. (a-b) Side and top views of the phantom with an optically-thin sheet of nanoparticles as a planar target. (c) Setup of the EIR measurement. The nanoparticle sheet was positioned perpendicular to the imaging plane of the ring-array transducer.

### D. Low-pass filter design with optimal cutoff frequency

To retain the integrity of the passband signal, we used a second-order low-pass Butterworth filter (LPF) because of its flat response in the passband [43]. The second-order LPF provided a damping ratio of 0.707 and had adequate roll-off in gain after the cutoff frequency while retaining a flat response in the passband. To choose the optimal cutoff frequency, the LPF was applied with cutoff frequencies ranging from 10 kHz to 7 MHz, and images were reconstructed with the delay-and-sum method [44]. The contrast (Eq. (6)) in the reconstructed region of interest (ROI) from each band is defined as

$$Contrast = 20\ log_{10}\left(\left|\frac{\mu_{ROI}}{\mu_{BGD}}\right|\right), \qquad (6)$$

where $\mu_{ROI}$ is the target mean signal and $\mu_{BGD}$ is the background mean signal [45]. A representative area of the image with functional or molecular information was chosen for each experiment. For example, in the photoswitching phantom experiment, the ROI included the cells. The cutoff frequency for the LPF was experimentally-dependent and corresponded to the peak contrast value for molecular and functional analysis. When no peak was found between 10 kHz and 7 MHz, an empirical cutoff frequency of 0.5 MHz was used. The LPF cutoff frequencies for each experiment are listed in Table II. Low-pass signals were compared with all-pass and band-pass signals. Cutoff frequencies of 2.5 MHz and 7.5 MHz were chosen for the band-pass filter (BPF), because this range covers the nominal bandwidth of the ring-array transducer used in PACT with 5-MHz center frequency at 100% bandwidth [39, 46]. Thus, it is worth noting that there could be a frequency gap between LPF and BPF, depending on the optimal cutoff frequency for the LPF.

TABLE I
LASER ENERGY USED IN EACH EXPERIMENT (TOTAL ENERGY FROM ALL FIBER OUTLETS).

| EXPERIMENT | Laser Energy (mJ) |
|---|---|
| Photothermal phantom (*in vitro*) | 36 (1064 nm) |
| Photoswitching cell phantom (*in vitro*) | 20 (750 nm) |
| Photoswitching mouse (*in vivo*) | 31.1 (750 nm) |
|  | 50.5 (1064 nm) |
| Hypoxia challenge (*in vivo*) | 8.8 (750 nm) |
|  | 20 (1064 nm) |

### E. Gold nanostar phantom

The nanoparticles used in our experiments were highly-absorbing gold nanostars (GNS). Increasing interest in nanoparticle-mediated thermal therapies [47, 48] has required the development of a biocompatible GNS. We succeeded in developing a novel surfactant-free synthesis method to make GNS biocompatible for *in vivo* applications [49, 50]. These GNS have a tunable plasmonic resonance within the therapeutic window of 750–1000 nm, making them promising photothermal agents for solid tumor ablation, photoimmunotherapy, and multimodal theranostic applications [51, 52]. GNS are also good absorbers for PA imaging experiments, with strong optical absorption in the NIR window (Fig. 3a).

Surfactant-free GNS were prepared using the method developed by Yuan *et al.* [53]. Briefly, 1 mL of 10 nM 12-nm gold seed solution was added to a rapidly stirring mixture of 100 mL of 0.25 mM $HAuCl_4$ and 100 μL of hydrochloric acid. 500 μL of 4 mM $AgNO_3$ and 500 μL of 0.1 M ascorbic acid were then immediately added to the mixing solution. The GNS solution was stabilized by adding SH-PEG$_{5000}$ (final concentration, 10 μM). The solution was centrifuged and resuspended at 10 nM before phantom fabrication. Extinction spectra were obtained using a dual-beam spectrophotometer (Shimadzu UV-3600; Shimadzu Corporation, Japan). The GNS solution exhibited peak absorption at 749 nm (Fig. 3a). To verify GNS formation, particle morphology was observed via transmission electron microscopy (TEM) (FEI Tecnai G² Twin) (Fig. 3b).

For all PA phantoms, 100 mL of deionized water was brought to a boil and 3 g of agarose powder was added slowly to avoid clumping. The agarose solution was poured into the phantom molds and was cooled to room temperature before removal from the molds. The concentrated GNS solution was then added to aliquots of boiled agarose solution to achieve a concentration of 0.5 nM in the desired regions of each phantom.

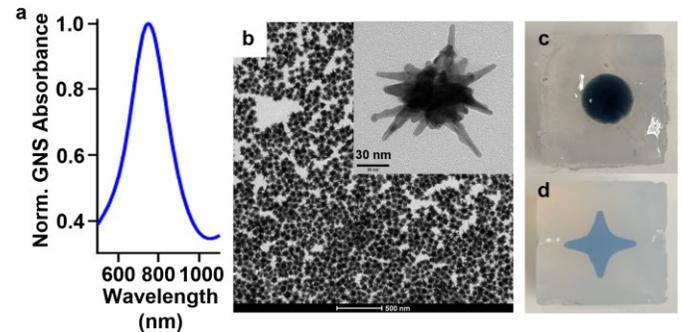

Fig. 3. GNS nanoparticles in gel phantoms for photothermal treatment and quantitative PA thermometry. (a) The normalized absorption spectrum of GNS. (b) TEM image of GNS; scale bar, 500 nm. Inset: the magnified image of a single GNS. (c) Spherical and (d) star-shaped GNS phantoms in clear agarose gel for PA imaging.

Several different GNS-based phantoms were used to mimic tissue models. (1) An ultra-thin layer of GNS sandwiched between two pieces of clear agarose gel was used to measure the EIR of the PACT system. A thin slab can generate a monopolar PA signal close to the Dirac delta function, whereas a point target generates a bipolar sawtooth waveform which is the derivative of the Dirac delta function [36, 40]. (2) A set of spherical GNS phantoms with a respective radius of 1, 3.3, 5, and 7 mm were embedded 2 mm beneath the agarose surface. Frequency spectra of the PA signals before and after EIR deconvolution from these spherical phantoms were compared with the analytical solution. (3) For PA temperature mapping in the photothermal experiment, a spherical phantom with a radius of 3.3 mm was used to mimic a tumor in SYMPHONY treatments (Fig. 3c) [52, 54]. A thermocouple (TC-08, Omega Engineering, Inc.) with its tip inside the phantom was used to validate the PA temperature measurements. A single-cycle treatment for 10 mins was used according to the standard photothermal therapy protocol with GNS [55, 56]. (4) To study the effect of sparse sampling, we used a star-shaped GNS phantom that had sharp spatial features (Fig. 3d).

### F. Photoswitching experiment with DrBphP-PCM-expressing tumor cells and BphP1-expressing mice

To examine the effect of low-frequency signals on photoswitching PA imaging *in vitro*, we used 4T1 mouse breast cancer cells that stably expressed DrBphP-PCM, a NIR photoswitchable phytochrome [57]. Plasmids co-encoding DrBphP-PCM and EGFP were used to transduce 4T1 cells via a lentivirus. The co-expressed EGFP signals were used to select stable DrBphP-PCM-expressing cells. Transduced 4T1 cells were cultured in DMEM with 10% FBS and 1% penicillin at 37°C in 5% $CO_2$ atmosphere. Approximately $5 \times 10^6$ 4T1 cells were washed three times with 50 mL PBS. The cells were concentrated at 1000 rpm for 5 minutes, then embedded inside an agarose phantom with a radius of 5 mm.

TABLE II
LPF CUTOFF FREQUENCY IN EACH EXPERIMENT, DETERMINED BY THE CONTRAST.

| EXPERIMENT | Cutoff Frequency (MHz) |
|---|---|
| GNS photothermal phantom (*in vitro*) |  |
|    Single-cycle treatment | 0.95 |
|    Three-cycle treatment | 1.95 |
| Photoswitching cell phantom (*in vitro*) | 3 |
| Photoswitching in mice (*in vivo*) | 0.25 |
| Hypoxia challenge (*in vivo*) | 0.51 |

For *in vivo* photoswitching experiments, BphP1-expressing mice were used following the procedure in Kasatkina *et al* [58]. Homozygous loxP-BphP1 mice (males, 6 months old) were anesthetized with 2% isoflurane through a face mask during the surgical procedure, and body temperature was maintained at 37.0 ± 0.2 ºC [58]. Mice were kept in the prone position. The lower back area was shaved and cleaned with iodine and alcohol. A 1 cm longitudinal skin incision was made below the rib edge at the location of 1 cm lateral to the midline. The muscle of the abdominal wall was cut using a high-temperature cautery loop tip. Two small retractors were placed, and the kidney was exposed. A 25G needle with a 5-µL Hamilton syringe was inserted 5 mm into the left kidney, and 1 µl of Cre-expressing AAV vector was injected slowly, while the right kidney was used as the control. The needle remained in position for 5 minutes and then was removed. The muscle and skin layers were sutured separately [58]. At 22 days after the AAV injection, the mice were imaged by PACT, using the photoswitching light illumination strategy described above. All animal procedures were approved by Institutional Animal Care and Use Committee (IACUC) at Duke University.

### G. Hypoxia challenge

A hypoxia challenge was performed with mice under anesthesia using 1% isoflurane at a 1.5 L/min flow rate. The experiment consisted of normoxia and hypoxia conditions. For normoxia conditions, the breathing air was a mixture of 21% oxygen and 79% nitrogen; for hypoxia conditions, the breathing air was 2% oxygen and 98% nitrogen. Each experiment started with 1 minute of normoxia baseline, followed by 30 s of hypoxia challenge, and 6 minutes of recovery under normoxia conditions. The hypoxia/normoxia cycle was repeated three times. Two wavelengths at 750 nm and 1064 nm were chosen, because deoxy- and oxy-hemoglobin were respectively the dominating absorbers at these wavelengths [59, 60]. All animal procedures were approved by IACUC at Duke University.

### H. PA image reconstruction and quantitative analysis

RF signals were first deconvolved with the measured EIR using Wiener deconvolution to improve low-frequency components [42]. Images are then reconstructed from the deconvolved RF data using the delay-and-sum method [44]. The speed of sound (SOS) was determined by measuring the coupling medium temperature during imaging. For photothermal experiments, the change in the phantom SOS was calibrated by measuring the arrival time of the PA signals. Reconstructed bipolar images were converted into unipolar images by multi-view Hilbert transform (MVHT) for *in vivo* data, and by thresholding negative pixels for phantom data [61].

For functional and molecular imaging experiments, additional quantitative analysis was applied. The relative temperature change during photothermal experiments was calculated from PA signals based on the relationship between the Grüneisen coefficient $\Gamma$ and PA signal amplitude $p_0$ as [62, 63]

$$p_0 = \Gamma \eta_{th} \mu_a F \quad (7)$$

where $\eta_{th}$ is the thermal conversion percentage, $\mu_a$ is the optical absorption coefficient, and $F$ is the optical fluence. In agar phantoms, $\Gamma$ has a linear dependence on temperature $T$ [64, 65]

$$\Gamma(T) = 0.04T + 0.11 \quad (8)$$

From Eqs. (7) and (8), the relationship between PA signal change $\Delta p_0$ and relative temperature change $\Delta T$ is [66]

$$\frac{\Delta p_0}{p_{0,0}} = \frac{0.04 \Delta T}{0.04 T_0 + 0.11} \quad (9)$$

where $T_0$ and $p_{0,0}$ are the baseline temperature and PA signal amplitude, respectively. Rewriting Eq. (9), we can estimate $\Delta T$ as

$$\Delta T = \frac{\Delta p_0}{p_{0,0}} (T_0 + 2.75) \quad (10)$$

To estimate $sO_2$ in the hypoxia challenge experiment, we applied linear spectral unmixing to post-MVHT images acquired at 750 nm ($P(\lambda_1, x, y)$) and 1064 nm ($P(\lambda_2, x, y)$) to estimate the molar concentration of HbR ($C_{HbR}$) and $HbO_2$ ($C_{HbO_2}$), assuming uniform optical fluence [5, 67]:

$$\begin{bmatrix} C_{HbR}(x,y) \\ C_{HbO_2}(x,y) \end{bmatrix} = (\varepsilon^T \varepsilon)^{-1} \varepsilon^T \begin{bmatrix} P(\lambda_1, x, y) \\ P(\lambda_2, x, y) \end{bmatrix} \quad (11)$$

in which

$$\varepsilon = \begin{bmatrix} \varepsilon_{HbR}(\lambda_1) & \varepsilon_{HbO_2}(\lambda_1) \\ \varepsilon_{HbR}(\lambda_2) & \varepsilon_{HbO_2}(\lambda_2) \end{bmatrix} \quad (12)$$

where $\varepsilon_{HbR}(\lambda_i)$ and $\varepsilon_{HbO_2}(\lambda_i)$ are the molar extinction coefficients of HbR and $HbO_2$ at two wavelengths, respectively. The signal processing pipeline is summarized in Fig. 4.

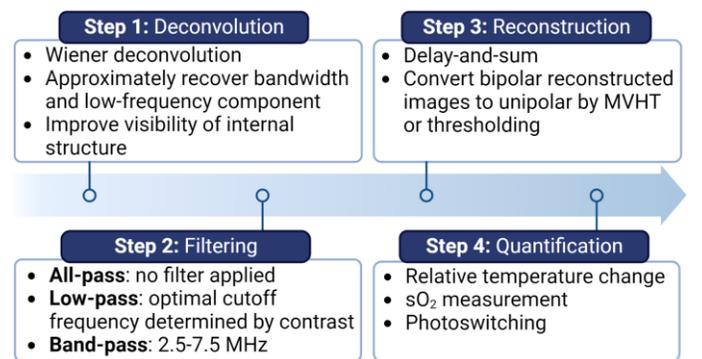

Fig. 4. Signal processing, imaging reconstruction, quantitative analysis pipeline.

### III. RESULTS

### A. System electrical impulse response

The receiving (one-way) EIR and transfer function of the system are shown in Fig. 5. The Fourier transform of the EIR signal shows a −6 dB receiving bandwidth of 0.15–7 MHz (Fig. 5a), which reflects the system's high sensitivity to low-frequency signals. The wide bandwidth allowed for quantitative comparison between different frequency bands in this study. By comparison, the −6 dB transmitting-receiving (two-way) bandwidth of the system is 3.5–6.5 MHz (Fig. 5a). The reconstructed image of the planar nanoparticle target is shown in Fig. 5b.

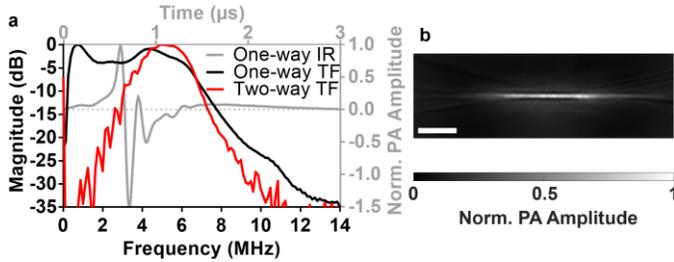

Fig. 5. System characterization using a planar target. (a) EIR and TF of the ring-array system, extracted from the one-way RF signal of the planar target. (b) Corresponding PA image of the planar target. Scale bar, 2 mm.

### B. PA signal frequency dependence on the target size

To validate the measured EIR of the PACT system, the frequency spectra of the deconvolved RF data and reconstructed PA images of spherical GNS phantoms were compared with the analytical solutions of spherical targets (Fig. 6a) [68]. The limited-band (LB) analytical solution is generated by convolving the full-band (FB) analytical solution with the measured EIR. As expected, the experimental images without deconvolution and the analytical images with LB solution had hollow inner structures (Fig. 6b, 6d). These hollow structures even appeared in LB analytical solution where optical fluence attenuation was not considered. Thus, these hollow structures were resultant from the limited bandwidth, but not from the optical fluence attenuation. The only exception was the 1-mm-radius sphere, simply because this small sphere did not generate strong low-frequency signals. Reconstructed experimental images with deconvolved signals of homogenous sphere structures were similar to the FB analytical solutions (Figs. 6c, 6e). The only exception was the 7-mm-radius sphere, on which the effect of optical fluence attenuation became significant. The improvement in internal structural visibility in Fig. 6c shows the importance of deconvolution as a necessary step before image reconstruction to boost the low-frequency components.

We further computed the structural similarity index (SSIM) and peak signal-to-noise ratio (PSNR) between the reconstructed experimental images and FB analytical solution, showing 2–3 fold higher SSIM and PSNR for the experimental images with deconvolution than those without the deconvolution (Figs. 6f-g). The SSIM and PSNR had no difference for the 1-mm-radius sphere because the small sphere's frequency spectrum could be fully covered by the system's detection bandwidth even without deconvolution. Interestingly, both SSIM and PSNR reached the maximum for the 3.3-mm-radius sphere with the deconvolved signals, suggesting that this may be the closet to the target size best matched with the system's bandwidth, as long as the optical fluence attenuation was negligible.

The image profiles in Figs 6b-e are shown in Fig. 7a. The frequency spectra shifted towards lower frequencies with the increasing target size in the deconvolved data and FB analytical solution, but not in the original data and LB analytical solution (Fig. 7b). The frequency shift shows that deconvolution using the measured EIR signal can broaden the bandwidth of the experimental data, particularly in the low-frequency range. The wide receiving bandwidth (0.15–7 MHz) of our customized transducer allows for comparing the effects of low-frequency components with the band-pass signals (2.5–7.5 MHz). The frequency shift in the deconvolved data also matched the analytical solution (Figs. 7b and 8a).

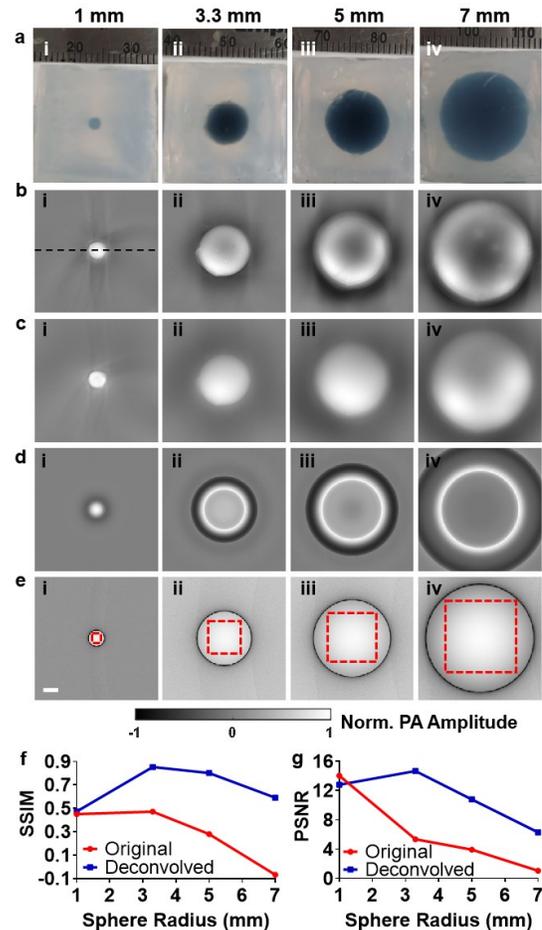

Fig. 6. Reconstructing sphere targets without deconvolution (original), with deconvolution, and analytical solution. (a) Photographs of the sphere phantoms. Reconstructed images of spheres with a radius of (i) 1 mm, (ii) 3.3 mm, (iii) 5 mm, and (iv) 7 mm using (b) experimental data without deconvolution, (c) experimental data with deconvolution, (d) band-pass analytical solution, and (e) all-pass analytical solution. Reconstructed images in (b-d) are normalized to the maximum absolute PA signal amplitude, while those in (e) are normalized between -1 and 1. (f) SSIM and (g) PSNR of the reconstructed experimental images with and without deconvolution compared to the full-band analytical solution. The red dashed rectangles indicate the ROIs for SSIM and PSNR calculation. The dashed line in (b) denotes the image profiles shown in Fig. 7. Scale bar, 2 mm.

The size-dependent frequency of the PA signals demonstrates the importance of low-frequency signals for detecting targets with large sizes. Figure 8a shows that frequency spectra of PA signals from 5-mm- and 7-mm-radius spheres decrease quickly down to 0.1 MHz. Thus, a detection system that cannot provide a frequency bandwidth below 1 MHz may only detect the boundaries of large organs with relatively homogeneous optical properties, such as the kidney, liver, and brain. Again, the frequency shift highlights the importance of deconvolution in

When the sphere radius increases from 1 mm to 5 mm, the peak frequency drops from 0.4 MHz to 0.1 MHz, while the signal magnitude increases by 20 dB. This result suggests a better detection sensitivity at low frequency is beneficial for detecting large targets, which has contributed to the improved functional and molecular imaging described below.

### C. In vitro temperature mapping at different frequency ranges

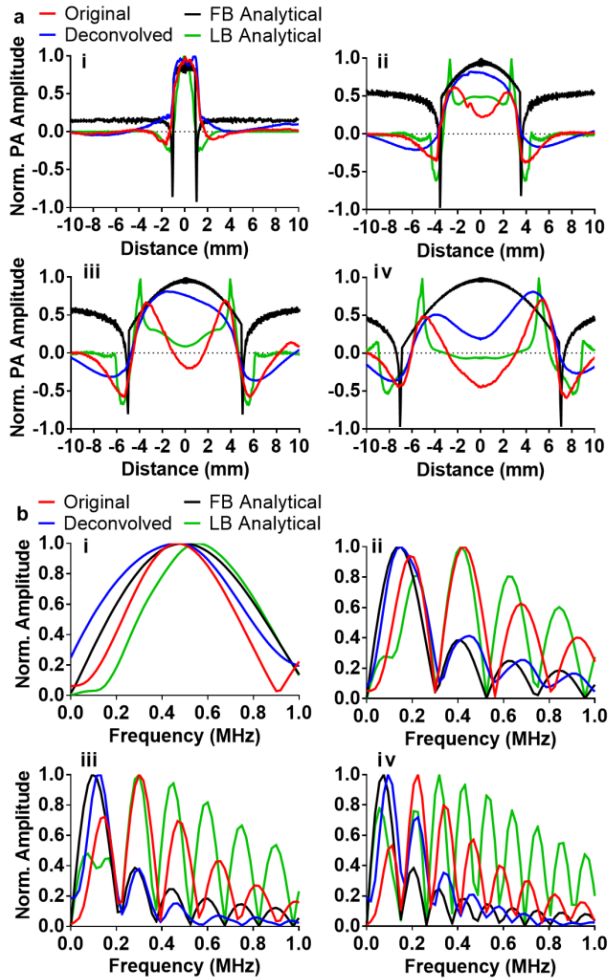

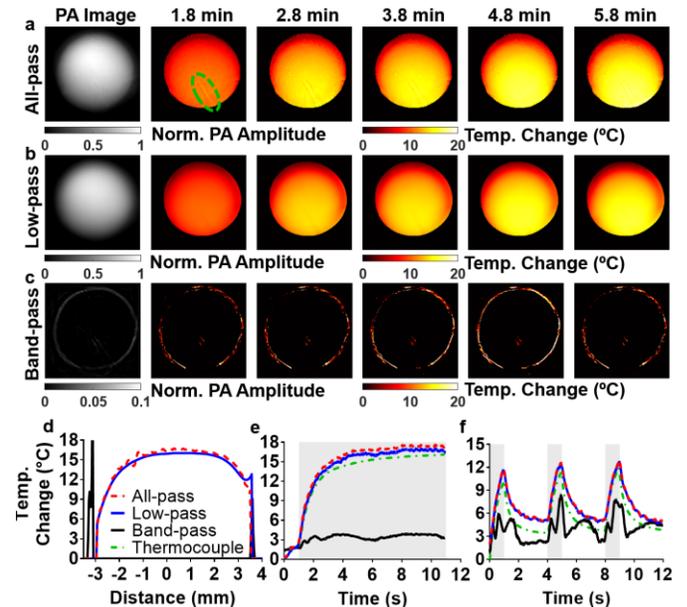

Fig. 9. Mapping relative temperature change during photothermal treatment. (a-c) Temperature maps at incremental timepoints in single-cycle photothermal treatment, with PA images of a sphere reconstructed from all-pass, low-pass, and high-band-pass (2.5–7.5 MHz) data respectively. The green ellipse in (a) at 1.8-min timepoint denotes the thermocouple region for signal averaging. (d) Spatial profile of temperature change along the red dashed line in (a). (e-f) Time courses of relative temperature change with single-cycle and three-cycle-treatment, respectively. Shaded regions indicate the time during which the CW laser was turned on. All temporal plots were smoothed with a moving average of 60 time points. Scale bar, 2 mm.

Fig. 7. Comparison of reconstructed image profiles and frequency spectra without deconvolution, with deconvolution, with LB and FB analytical solution. (a) PA image profiles along the dashed line in Fig. 6b-i with a sphere radius of (i) 1 mm, (ii) 3.3 mm, (iii) 5 mm, and (iv) 7 mm in Fig. 6. (b) The corresponding signal frequency spectra with different sphere sizes.

recovering the inner structures of large targets (Fig. 8a).

Further investigation on the size-dependent PA signal using FB analytical solution shows the increased PA signal amplitude and the decreased peak frequency with the target size (Fig. 8b).

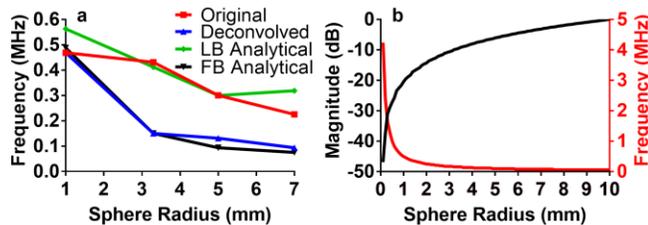

Fig. 8. Size-dependent peak frequency shift and signal magnitude from PA signals of the sphere phantoms and analytical solution. (a) Peak signal frequency versus the sphere radius. (b) PA signal magnitude and peak frequency versus the sphere radius, using the FB analytical solution.

Temperature mapping for photothermal treatment of GNS spheres using all-pass, low-pass, and band-pass (2.5-7.5 MHz) data are shown in Fig. 9. Qualitatively, reconstructed images from all-pass and low-pass data have comparable visibility of the sphere and similar temperature maps over time (Figs. 9a, 9b). We can observe a thermal gradient from the hotter core towards the cooler boundary of the sphere (Fig. 9d). In contrast, the frequency band of 2.5–7.5 MHz only reconstructs the edge of the sphere, and the center of the sphere was not visible (Figs. 9c, 9d). For all-pass and low-pass data, the PA-measured temperature around the thermocouple tip correlated well with readings from the thermocouple for both single and three heating cycles (Figs. 9e, 9f). Temperature quantification from the low-pass data has better accuracy for the single-cycle heating than the all-pass data. Temperature measurement with the band-pass data shows substantial underestimation, especially in single-cycle heating (Fig. 9e). Thus, using low-frequency signals, quantitative temperature mapping can be more reliable.

### D. In vitro photoswitching imaging of cells expressing DrBphP-PCM

As an example of molecular imaging, PACT images of photoswitching cells expressing DrBphP-PCM using different frequency ranges are shown in Fig. 10. By subtracting the OFF-state image from the ON-state image of the cells, the resultant differential image has enhanced image contrast due to the reduced non-switching background signals [58]. Differential images of the DrBphP-PCM-expressing cells (Figs. 10a and 10c) show clear photoswitching signal changes for both all-pass and low-pass signals. However, using the band-pass signals, we cannot observe clear photoswitching signals from the cells. Due to the relatively homogeneous distribution of cells as a bulk PA target, PA signal from the cell phantom has a low-frequency band peaking at 0.16 MHz with a narrow bandwidth (Fig. 10d). Thus, a band-pass filter of 2.5–7.5 MHz removed almost all the signals from the cells (Fig. 10e). Figure 10(f) shows four repeated photoswitching cycles using low-frequency signals with an average switching ratio of 25% (Fig. 10g). In contrast, the band-pass data failed to capture the photoswitching process, showing only transient PA signal increase due to laser heating.

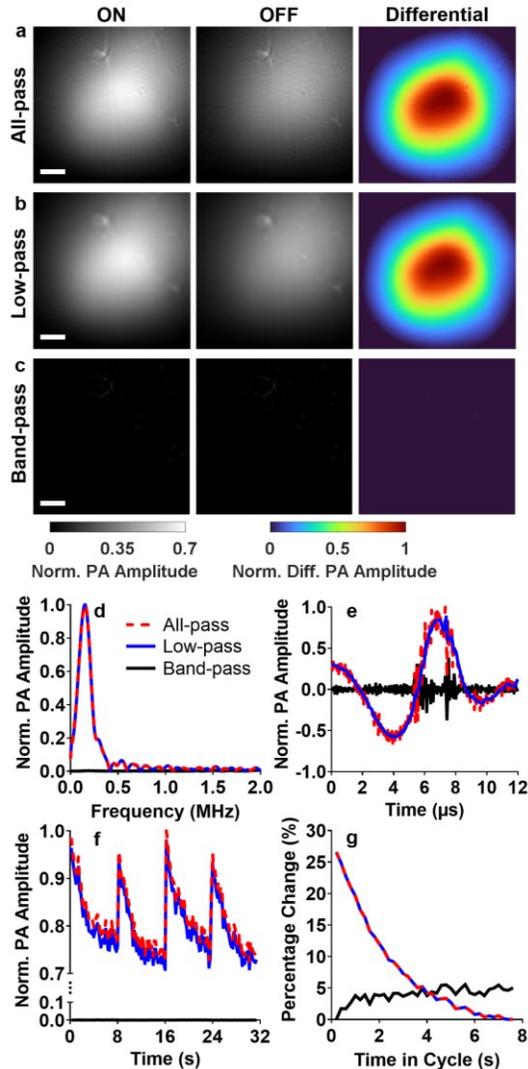

Fig. 10. Detection of photoswitching phytochromes on a cell phantom. (a-c) ON-state, OFF-state, and differential PA images of the cell phantom from (a) all-pass, (b) low-pass and (c) band-pass data. (d) Frequency spectra and (e) RF signals of the cell phantom with different frequency ranges. (f) Time-course of the averaged PA signal amplitude of the cell phantom over four photoswitching cycles. (g) The averaged PA signal changes over four cycles. Scale bar, 1 mm.

### E. In vivo photoswitching imaging of a mouse kidney expressing BphP1

The whole-body image of a mouse kidney region is shown in Fig. 11a. The control (right) kidney exhibits no photoswitching compared to the left kidney (Fig. 11b). Here, in vivo quantification of photoswitching kidney tissue with low-frequency signals shows clear improvement over the all-pass and band-pass signals (Fig. 11c). While the differential image with the all-pass data shows a photoswitching area of 19.6 mm$^2$ near the skin surface, the differential image with the low-pass data shows photoswitching signals at depths up to 8 mm from the skin surface, with a total photoswitching area of 61.6 mm$^2$ (Figs. 11d, 11e). The differential image with the band-pass signals shows little photoswitching signals. The BphP1-expressing kidney had a relatively uniform expression level, resulting in the differential signals concentrated largely in the low-frequency range. In addition, it is clear that low-frequency signals have greater photoswitching amplitude, enabling deeper and more sensitive PA detection of the molecular probes. One-way ANOVA test for multiple comparisons presents a significant increase in the photoswitching amplitude when using the low-pass data ($p < 0.0001$) (Fig. 11f).

In addition, it is evident that the band-pass signals were more sensitive to the breathing motions, which decreased the cross-correlation over the consecutive frames (Fig. 11g). An unpaired t-test on cross-correlation of breathing frames shows a significant difference between the all-pass and low-pass data ($p < 0.0001$) (Fig. 11h), mostly because the all-pass data contains the motion-sensitive band-pass data. Since the photoswitching quantification depends on averaging multiple switching cycles over an extended period of time (176 seconds for 11 switching cycles), the improvement of low-pass data over all-pass data is due to suppression of the band-pass signal fluctuations induced by breathing motion. By contrast, under the motionless circumstances in the phantom experiment in Section III D, all-pass and low-pass data had similar quantification performance.

### F. In vivo mapping of blood oxygenation level in hypoxia challenge

Blood oxygenation (sO$_2$) measurement is a common application of PA imaging. Here, we studied the impact of low-frequency signals on whole-body temporal and spatial sO$_2$ estimation in the hypoxia challenge. Figures 12a and 12b are PA images of the mouse liver region under normoxia and hypoxia at 1064 nm and 750 nm. Temporally, low-pass signals yielded less sO$_2$ fluctuations with a standard deviation (SD) of 8.8% over the baseline measurement, compared to 12% and 25% from all-pass and band-pass signals, respectively (Fig. 12f). The low-pass signals also yielded more homogeneous sO$_2$ mapping in space over three cycles of the hypoxia challenge (Figs. 12d, 12f). Spatial profiles reveal improvement in the stability of sO$_2$ measurements from low-pass signals, with the spatial SD below 6% throughout the mouse cross-section both under normoxia and hypoxia (Figs. 12g, 12h). In contrast, because high-frequency signals were more attenuated with increasing depth, both all-pass and band-pass signals led to noisier $sO_2$ measurements, especially at deeper regions. $sO_2$ quantification was less accurate in hypoxia, due to the reduced PA signal amplitudes, especially at 1064 nm where the absorption of oxy-hemoglobin is dominant. For all-pass signals, the spatial SDs increased from ~7.9% in normoxia to ~10% in hypoxia (Figs. 12g, 12h). For band-pass signals, $sO_2$ measurements were substantially less accurate than the low-pass and all-pass signals, especially under hypoxia (Fig. 12h). At all depths, band-pass results showed strong fluctuations with spatial SDs of ~20% under normoxia and ~28% under hypoxia (Figs. 12g, 12h). The improvement in $sO_2$ measurements using low-pass signals is useful for tracking hemodynamic changes in deep tissues.

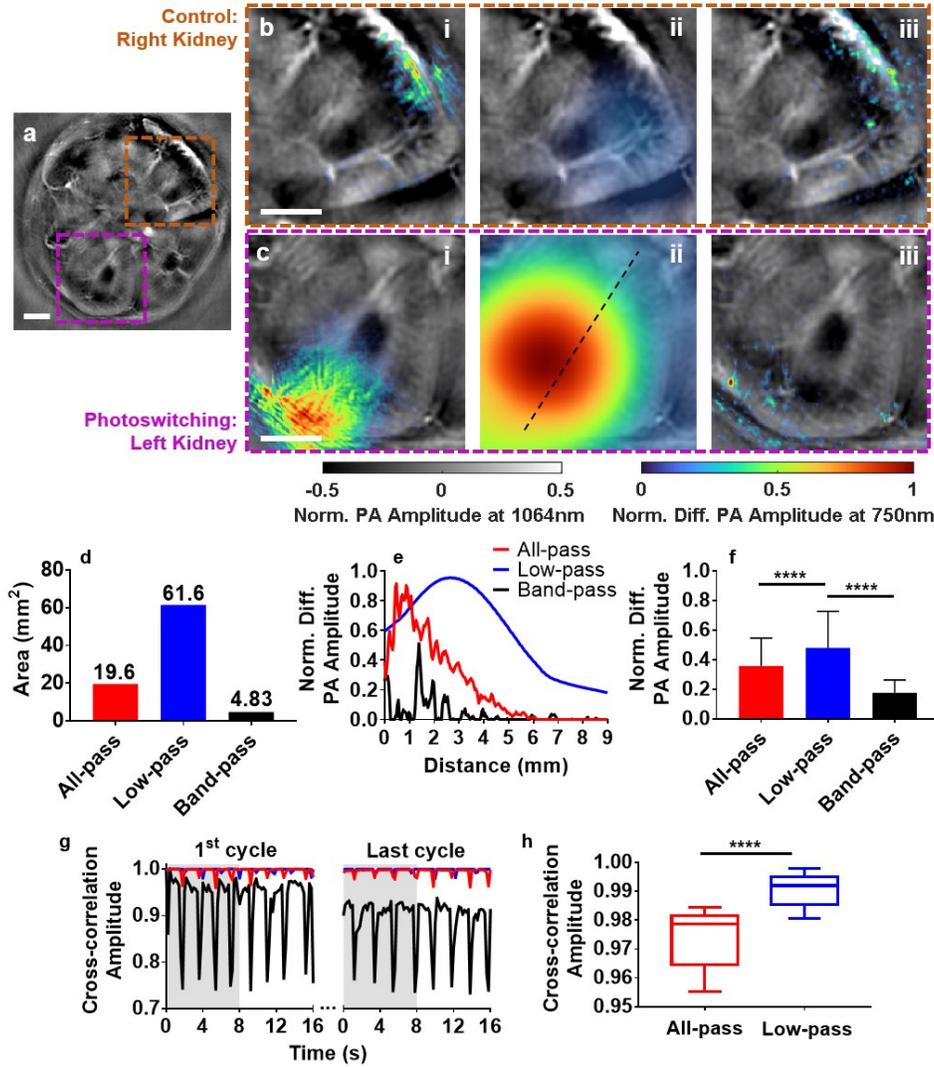

Fig. 11. *In vivo* photoswitching imaging of mouse kidney. (a) The whole-body cross-sectional PA image at 1064 nm around the kidney region. (b-c) Close-up PA images of the right (control) and left (photoswitching) kidney, overlaid with the differential image at 750 nm reconstructed with (i) all-pass, (ii) low-pass, and (iii) band-pass signals. (d) Photoswitching area quantified from differential images in (c). (e) Spatial profiles of the reconstructed differential images along the dashed line in (c)-ii. (f) Averaged differential PA signal amplitudes. (g) Cross-correlation of the PA signals over two photoswitching cycles, showing the band-pass signals are more sensitive to the breathing motions. The shaded regions indicate turning on the phytochromes. (h) Cross-correlation amplitude of breathing frames with the all-pass and low-pass data. **** $p < 0.0001$. Scale bar, 2 mm.

### G. Imaging complex targets with sparse-sampling

Sparse sampling is a common source of artifacts in PACT and is often caused by the lack of sufficient transducer elements. Here we examined the impact of low-frequency signals on sparse-sampling artifacts by imaging star-shaped phantoms that have sharp features and complex spatial-frequency patterns. We artificially induced sparse sampling by reducing the number of equally-spaced transducer elements instead of using the full aperture of 512 elements. A LPF cutoff frequency of 0.5 MHz was empirically selected. We observed increasing streaking artifacts with decreasing number of transducer elements, using all-pass and band-pass signals (Figs. 13a, 13c). In contrast, low-pass signals yielded reduced streaking artifacts with fewer transducer elements compared to all-pass and band-pass signals (Fig. 13b). In addition, we calculated SSIM and PSNR with sparse sampling compared with the full detection aperture (i.e., 512 transducer elements). Relatively consistent imaging quality was maintained by using only 32 or 64 elements with low-frequency signals, which outperformed both all-pass and band-pass signals (Figs. 13d, 13e). Thus, reconstructing PACT images with low-frequency signals can reduce the number of elements required for spatial anti-aliasing, particularly for large targets.

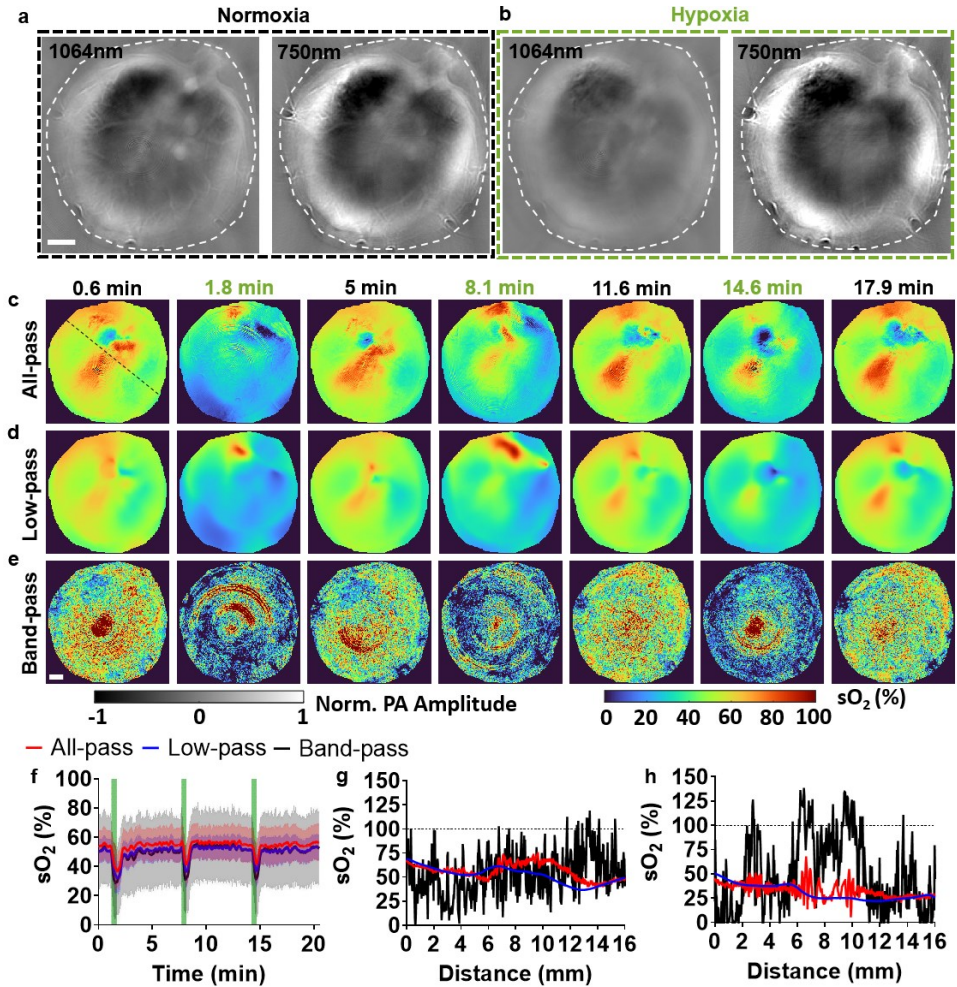

Fig. 12. Whole-body cross-sectional blood oxygenation measurement in hypoxia challenge. (a-b) Representative bipolar PA images of the mouse liver region under (a) normoxia and (b) hypoxia, acquired at 1064 nm and 750 nm. White dashed lines denote the mask to separate the mouse body from background for $sO_2$ quantification. (c-e) $sO_2$ mapping under normoxia and hypoxia conditions using (c) all-pass, (d) low-pass, and (e) band-pass data. (f) Temporal profile of averaged $sO_2$ in the whole cross-section, smoothed with a moving average of 60 timepoints. Shaded regions are the standard deviations of $sO_2$ of the whole cross-section at each timepoint. The green boxes denote hypoxia challenge. (g-h) Spatial profiles along the dashed line in (c) with the animal under (g) normoxia and (h) hypoxia, respectively. Scale bar, 2 mm.

## IV. Discussion

In this study, we investigated the importance of low-frequency signal components in functional and molecular PACT. We showed that low-frequency signals can improve the quantitative accuracy of functional and molecular applications, especially for relatively large and homogeneous targets. Guided by the PA signal spectrum from analytical solutions, we experimentally verified the dependence of the PA signal frequency on the characteristic size of the target. This dependence strongly suggests that large PA targets with relatively homogenous optical and acoustic properties, such as the liver and kidney, can generate strong low-frequency signals. We demonstrated the importance of low-frequency signals in various functional and molecular applications by using *in vitro* phantoms made of GNS nanoparticles, photoswitching cells, and *in vivo* animal models expressing photoswitching phytochromes or under hypoxia challenge. The experimental results confirmed that low-frequency signals are beneficial for imaging large structures and quantifying temperature, blood oxygenation, and photoswitching molecular probes.

In traditional PACT, low-frequency signals are usually omitted either due to the detection system or signal filtering. This lack of low-frequency signals is at least partially responsible for the low visibility of inner structures of large targets, while the optical fluence attenuation also contributes to the weaker inner structures. Such deterioration of structural integrity hinders accurate quantitation of functional and molecular information. For example, in temperature mapping, band-pass signals resulted in hollow GNS spheres in reconstructed images, and temperature changes inside the phantom during photothermal treatment could not be measured. In contrast, low-pass signals correctly tracked temperature increases that agreed with thermocouple readings. In molecular imaging, band-pass signals were unable to capture clear photoswitching signals of the BphP1-expressing kidney, which were captured by low-frequency signals. Similar to the *in vivo* photoswitching experiment, in the hypoxia challenge experiment, we observed more consistent $sO_2$ quantification temporally and spatially when using low-pass signals for image reconstruction. This was especially evident in deeper regions. Such improvement in blood oxygenation measurements without optical fluence compensation is encouraging for deep-tissue applications such as functional brain imaging.

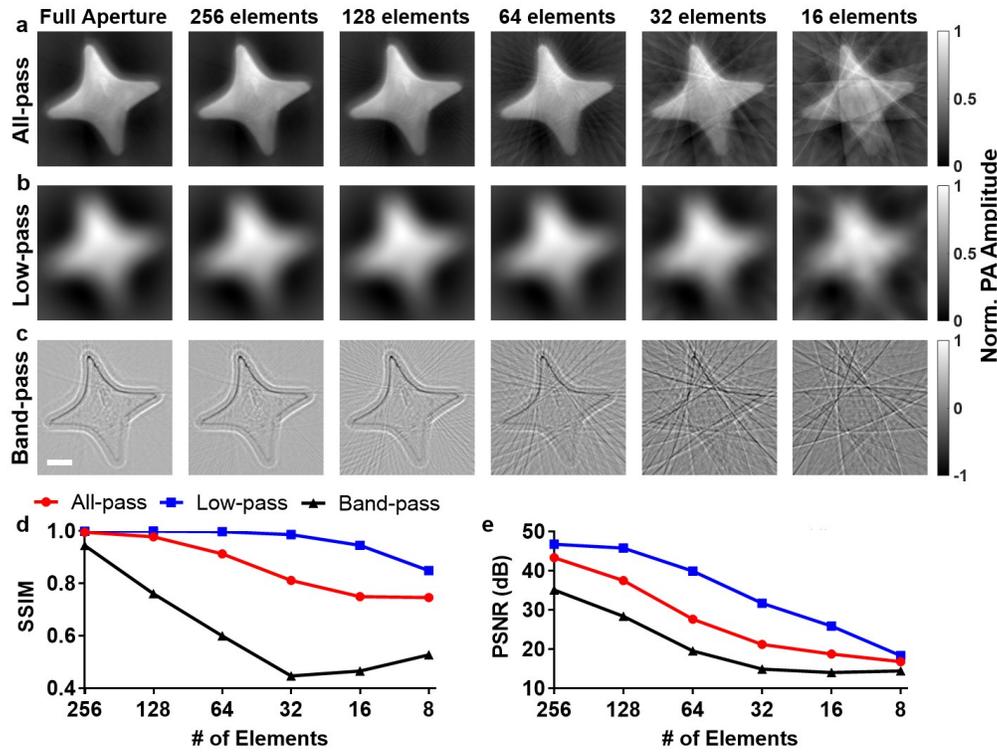

Fig. 13. Effects of sparse-sampling artifacts. (a-c) Reconstructed images with decreasing number of elements using all-pass, low-pass, and band-pass signals, respectively. (d-e) SSIM and PSNR of the sparse-sampling results compared with the full aperture detection. Scale bar, 2 mm.

Thus, in PACT, for large PA targets with relatively homogenous optical and acoustic properties, it is useful to utilize low-frequency signals to accurately quantify functional and molecular information. In practice, biodistributions of drugs and contrast agents inside tissues may be more homogenous than internal structures such as blood vessels, which favors the low-frequency signal components [69, 70]. Thus, for large organs with uniformly distributed molecular probes, low-frequency signals should be carefully preserved for functional and molecular data analysis.

Separating low-frequency signals from the all-pass signals is beneficial for *in vivo* applications, in which imaging quality is prone to adverse conditions, especially motion artifacts. For example, when imaging a BphP1-expressing mouse kidney *in vivo*, we showed that low-frequency signals improved the detection sensitivity when compared with all-pass and band-pass signals. Several factors contribute to this improvement: (1) the increased low-frequency signal magnitude of large PA targets, and (2) the suppression of high-frequency signal fluctuations due to motion artifacts. However, it is important to mention that these improvements in quantification do not apply to phantom experiments. In phantom experiments, motion artifacts are absent, and the low-frequency signals of homogenous targets already dominate. It is also crucial to acknowledge that while low-pass data can initially have a higher SNR than the band-pass data for large targets, the two bands may have comparable contributions for quantitative analysis when signal changes or ratiometric measurements are utilized.

We also observed that using low-frequency signals can reduce the impact of sparse-sampling artifacts. Compared to reconstructed images from all-pass and band-pass data, reconstructions from low-pass PA signals can be performed using 8–16 times fewer transducer elements while maintaining the image quality of complex targets. Collectively, our experimental results indicate that when PACT is used for functional and molecular applications and high spatial resolution is less critical, it is more robust to use low-frequency signals to achieve high quantitative accuracy and high detection sensitivity.

It should also be noted that using low-frequency signals in PACT has a key drawback of degraded spatial resolution, which may not be acceptable for biomedical applications that need high resolutions, such as visualizing small vessels, microbubbles, and nanodroplets [14, 15]. However, if the targets of interest have relatively large sizes and uniform optical and acoustic properties, trading resolution for quantitation accuracy may favor low-frequency signals. Another drawback of low-frequency signals is the artifact from transducer-light interaction when the initial SNR is low [71], which could be a source of error for quantitative analysis. Coating the transducer surface with optically reflective material may minimize the transducer-light interaction artifacts [71].

The low-frequency PACT can be improved in several aspects. First, better methods are needed for identifying the optimal cutoff frequency. Current contrast-based techniques are target-oriented and require prior knowledge of the ROI, such as the size of photoswitching organs. For applications in which target sizes are unknown, it is challenging to choose the optimal cutoff frequency. Second, further investigation is needed on different image reconstruction methods, improved deconvolution approaches, and varied detection geometries. For example, iterative Richardson-Lucy or regularized Tikhonov deconvolution can potentially improve low-frequency components over the Wiener deconvolution. Low-frequency components might also have different effects on PACT systems using a linear-array transducer with a severely limited view. Third, incorporating optical fluence compensation with low-frequency PACT can potentially improve quantification at deeper locations. Fourth, it is also interesting to study the low-frequency signals on the boundary build-up effect in PACT, particularly with a linear-array transducer. Improving the low-frequency signal detection may help mitigate the boundary build-up effect since the inner structures of large targets are dominated by low-frequency signals [72]. Last, another future direction is studying the effects of low-frequency signals on human imaging with larger and deeper organs. The advantages of low-frequency signals in functional imaging in humans such as early tumor detection and drug uptake in the brain will likely accelerate the clinical translation of PACT technologies.

## V. CONCLUSION

In summary, this study highlights the use of low-frequency signals in PACT to improve quantitative accuracy in functional and molecular imaging applications such as temperature mapping, blood oxygenation measurement, and molecular probe detection. We expect that our results will contribute to a better understanding of low-frequency signals in functional and molecular imaging and in optimizing PACT systems.